\documentclass[useAMS,usenatbib]{mn2e}

\usepackage{aas_macros}
\usepackage{graphicx}

\def\gtrsim{\mathrel{\hbox{\rlap{\hbox{\lower4pt\hbox{$\sim$}}}\hbox{$>$}}}}
\def\lesssim{\mathrel{\hbox{\rlap{\hbox{\lower4pt\hbox{$\sim$}}}\hbox{$<$}}}}

\title[Progenitor of SN~2013dk]{On the progenitor of the Type Ic  SN~2013dk in the Antennae Galaxies\thanks{ Based on observations collected at the European Organisation for Astronomical Research in the Southern Hemisphere, Chile; Program 091.D-0561.}
}
\author[Elias-Rosa et al.]{N. Elias-Rosa$^{1}$\thanks{E-mail: nelias@ieec.cat}, 
A. Pastorello$^{2}$, 
J.~R. Maund$^{3,4}$,
K. Tak\'ats$^{5}$,
M. Fraser$^{3}$,
S.~J. Smartt$^{3}$,
\newauthor S. Benetti$^{2}$,  
G. Pignata$^{5}$,
D. Sand$^{6}$ and
S. Valenti$^{7,8}$.  
 \\
 \\
$^{1}$Institut de Ci\`encies de l'Espai (IEEC-CSIC), Facultat de Ci\`encies, Campus UAB, 08193 Bellaterra, Spain.\\
$^{2}$INAF - Osservatorio Astronomico di Padova, vicolo dellÕOsservatorio 5, 35122 Padova, Italy. \\
$^{3}$Astrophysics Research Centre, School of Mathematics and Physics, Queen's University Belfast, Belfast BT7 1NN, UK.\\
$^{4}$Royal Society Research Fellow.\\
$^{5}$Departamento de Ciencias Fisicas, Universidad Andres Bello, Avda. Republica 252, Santiago, Chile. \\
$^{6}$Department of Physics, Texas Tech University, Lubbock, TX 79409, USA.\\
$^{7}$Las Cumbres Observatory Global Telescope Network, 6740 Cortona Dr., Suite 102 Goleta, Ca 93117, USA.\\
$^{8}$Department of Physics, University of California, Santa Barbara, CA 93106, USA.
}
\begin{document}

\date{Accepted .... Received ...; in original form ...\today}

\pagerange{\pageref{firstpage}--\pageref{lastpage}} \pubyear{2012}

\maketitle

\label{firstpage}

\begin{abstract}

We report the results of our search for the progenitor candidate of SN~2013dk, a Type Ic supernova (SN) that exploded in the Antennae Galaxy system. We compare pre-explosion {\sl Hubble Space Telescope\/} ({\sl HST\/}) archival images with SN images obtained using adaptive optics at the ESO Very Large Telescope. We isolate the SN position to within $3\sigma$ uncertainty radius of $0{\farcs}02$, and show that there is no detectable point source in any of the {\sl HST\/} filter images within the error circle. We set an upper limit to the absolute magnitude of the progenitor to be $M_{F555W}\gtrsim-5.7$, which does not allow Wolf-Rayet (WR) star progenitors to be ruled out. A bright source appears $0{\farcs}17$ away, which is either a single bright supergiant or compact cluster, given its absolute magnitude of $M_{F555W}=-9.02\pm0.28$ extended wings and complex environment. However, even if this is a cluster, the spatial displacement of SN~2013dk means that its membership is not assured. The strongest statement we can make is that in the immediate environment of SN~2013dk (within 10 pc or so) we find no clear evidence of either a point source coincident with the SN or a young stellar cluster that  could host a massive WR progenitor. 
\end{abstract}


%
%
\section[]{Introduction}

Massive stars evolve to an end state that results in the collapse of the stellar core, as hydrostatic pressure no longer balances gravity, producing a core-collapse supernova (CC-SN). These objects show a great diversity in their kinetic energy, luminosity and elemental abundances. This observational heterogeneity is chiefly related to the evolutionary state of the progenitor star at the moment of the explosion. Despite the large number of supernovae (SNe) discovered so far, the link between the progenitor evolution and the consequent explosion is not fully understood. Direct identification of the progenitor stars prior to the explosion have provided the most stringent constraints; we now know with an increasing degree of confidence that single red supergiants in the range of $\sim 8$--$16\ {\rm M}_{\sun}$ explode as Type II-plateau (II-P) SNe. We have also evidence that the progenitors of a few Type II-narrow (IIn) SNe are massive luminous blue variables, and that those of Type II-linear (II-L) and IIb SNe, are likely yellow supergiants (see e.g. \citealt{turatto07} for a discussion of SN classification, and \citealt{smartt_rev09} - and references therein - for a SN progenitor review). However, the origin of the stripped-envelope, hydrogen (or hydrogen and helium) poor Type Ib/c SNe is still debated. For several decades, these have been proposed to originate from either explosion of massive WR stars \citep{gaskell86}, or from lower mass helium stars in interacting binaries \citep{podsiadlowski92}. If they are massive, classical WR progenitors, then the absolute magnitude distribution of this system means that we may be able to detect them in pre-explosion imaging \cite[e.g. see ][for the deepest limits set]{crockett07}. Nevertheless, there has been no direct detection of their progenitors to date.
This is perhaps further evidence that the most massive stars may form quenched, black-hole forming SNe, while the majority of the Ib/c SNe we see come from lower mass helium stars in interacting binaries (e.g. see \citealt{eldridge13}). 

Here we present the case of SN~2013dk in NGC~4038, which is part of the well known Antennae galaxy system. SN~2013dk was discovered by the CHASE survey \citep{pignata09} on 2013 June 22.12 (UT dates are used throughout), and was soon after spectroscopically classified as a young stripped-envelope SN (\citealt{carrasco13}, \citealt{harutyunyan13}). The position of SN~2013dk was given as $\alpha = 12^{\rm h} 01^{\rm m} 52{\fs}72\, \pm 0{\fdg}2$, $\delta = -18\degr 52\arcmin 18{\farcs}3\, \pm 0{\fdg}2$ \citep[J2000.0;][]{carrasco13}. A first attempt at identifying the SN location was performed using archival\footnote[9]{\it http://archive.stsci.edu/hst/} {\sl HST\/} images from 2008-2009, with a reference image of the SN obtained at Telescopio Nazionale Galileo (TNG+Dolores) soon after the discovery \citep{eliasrosa13}. Here we present a more detailed study of the SN~2013dk progenitor candidate via further comparison of the {\sl HST\/} images with Very Large Telescope (VLT)+Naos-Conica (NaCo) adaptive optics data. 

%
%
\vspace{-0.5cm}
\section{The spectral type classification of SN 2013dk} \label{class}
\begin{figure}
\centering
{\includegraphics[width=0.55\columnwidth,angle=-90]{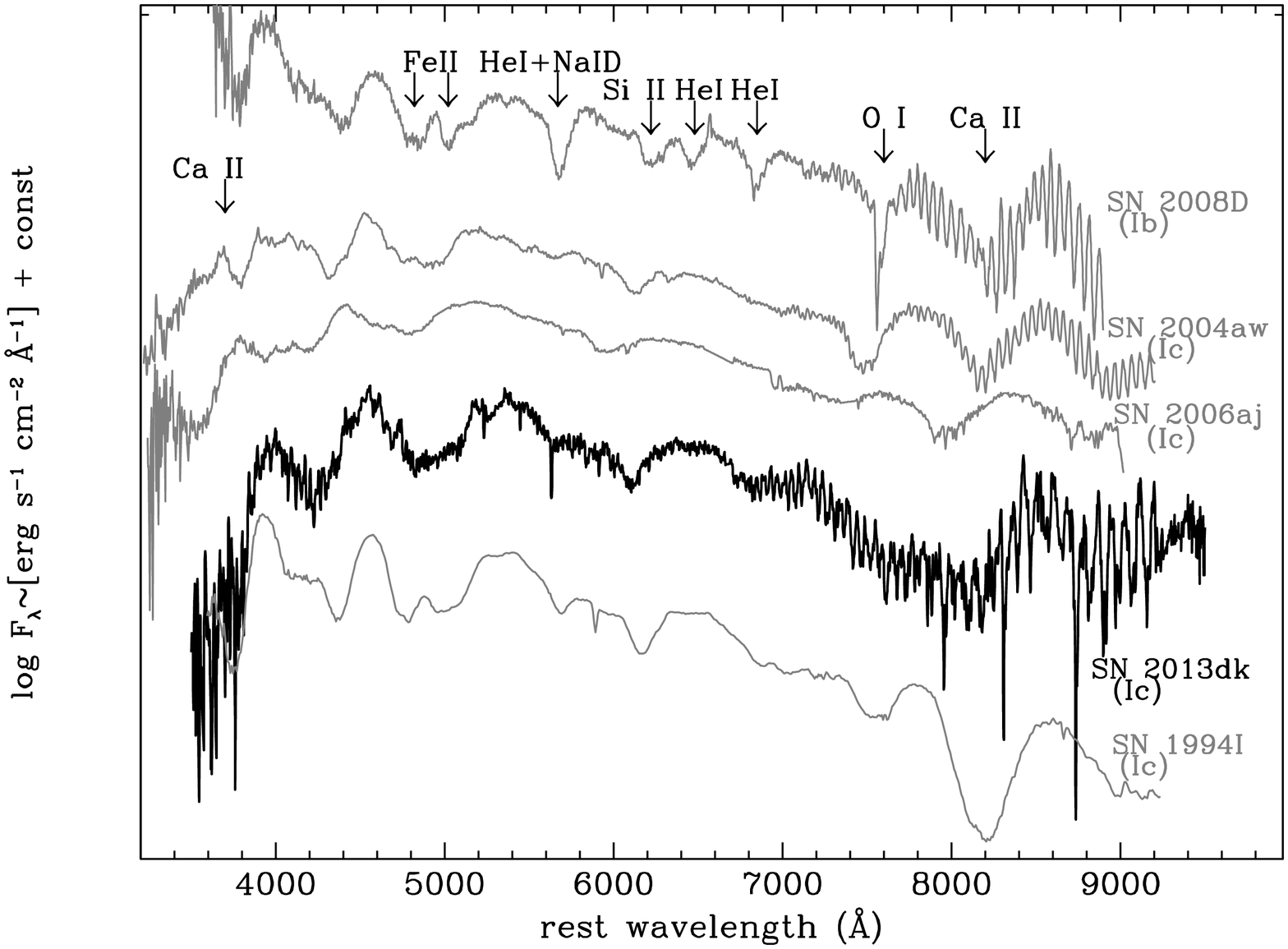}
\includegraphics[width=0.55\columnwidth,angle=-90]{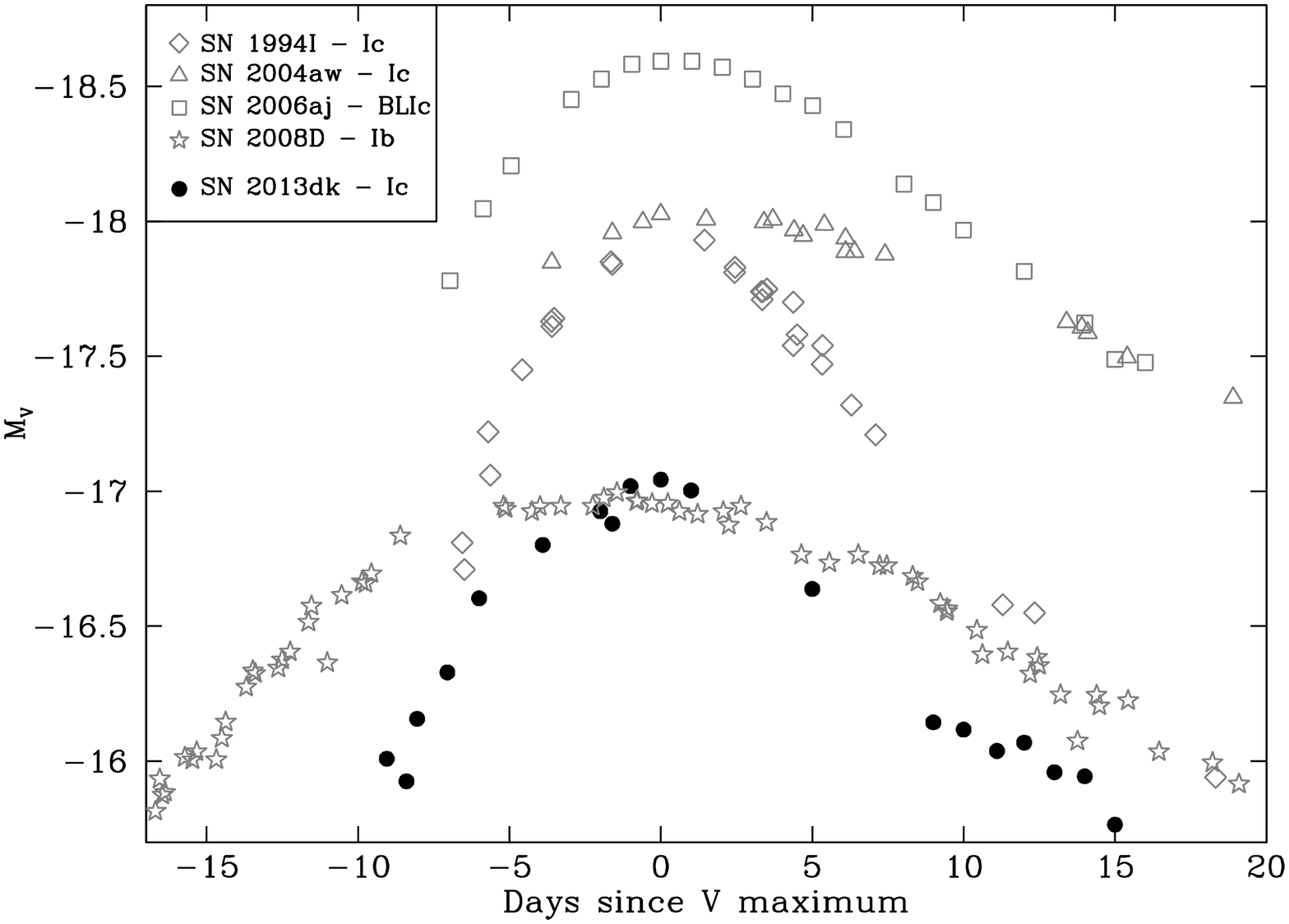}}
\caption{{\bf Top:} Comparison of reddening and redshift corrected spectra of stripped-envelope SNe at around the maximum light: SNe~2013dk, 1994I \citep{clocchiatti96}, 2004aw \citep{taubenberger06}, 2006aj \citep{pian06}, and 2008D \citep{malesani09}. {\bf Bottom:}  Comparison of the early-time and preliminary V-band absolute light curve of SN~2013dk with those of the Type Ic SNe~1994I \citep{richmond96}, 2004aw \citep{taubenberger06}, the broad-lined 2006aj \citep{pian06}, and the Type Ib SN~2008D \citep{mazzali08}.  The adopted distance moduli and total interstellar reddening values for the comparison objects were taken from the literature. }
\label{fig_sne}
\end{figure}

The classification spectrum reported in \citet{harutyunyan13} suggested that SN 2013dk was a Type Ic SN discovered nearly one week before maximum light. However, some stripped-envelope SNe develop weak He\,{\sc i} lines at later stages, and distinguishing between Ib and Ic SNe can be difficult. Soon after the announcement, our team started a multi-wavelength monitoring campaign for SN~2013dk, and a detailed analysis of the data obtained will be presented in a forthcoming paper (Tak\'ats et al. in prep.). Here, in order to establish a secure classification of SN~2013dk, we compare in Figure \ref{fig_sne} (top) a spectrum obtained on 2013 July 02.36 UT (close to maximum light) with the Faulkes Telescope South equipped with FLOYDS (\citealt{brown13}), to spectra of a sample of Type Ib/c SNe at a similar phase. The spectra of SN 2013dk, and the three Type Ic SNe do not show the evident He\,{\sc i} features that are clearly detected in SN~2008D, confirming the initial classification of SN~2013dk as a Type Ic event.

Assuming $E(B-V)_G = 0.04 \pm 0.01$ mag as the Milky Way component to the line-of-sight reddening \citep{schlafly11}, and measuring the $E(B-V)_{host}$ through the equivalent width (EW) of the Na\,{\sc i}~D line at the host-galaxy redshift ($z_0 = 0.0055$\footnote[10]{NED, NASA/IPAC Extragalactic Database; {\it http://nedwww.ipac.caltech.edu/}.}; \citealt{poznanski12}) in a set of low-resolution early spectra (Tak\'ats et al. 2013 in prep.), we derive a $E(B-V)_{\rm tot} = 0.24 \pm 0.02$ mag which will be adopted hereafter. However, we caution that due to large scatter seen for SNe in the EW(Na\,{\sc i}~D) vs.~$E(B-V)$ plane (e.g., \citealt{eliasrosa07} or \citealt{poznanski11}), the derived extinction should not be considered secure. Indeed, our estimate of the extinction could be underestimated, as suggested by the red colour of SN~2013dk. A detailed analysis of the extinction of this SN will be done after the completion of its monitoring campaign (Tak\'ats et al. 2013 in prep.). We adopt a distance of 22.3 $\pm$ 2.8 Mpc from Schweizer et al. (2008). The V-band absolute light curve of SN~2013dk (at maximum, on JD=2456476.5, M$_V$ = -17.02 mag) is compared in Figure \ref{fig_sne} (bottom) with those of a sample of type Ib/c SNe. We immediately note an overall similarity with the light curve of the broad-lined Ic SN~2006aj, although SN~2013dk shows a slightly faster post-maximum decline.  


%
%
\vspace{-0.5cm}
\section{Searching the progenitor candidate at the SN location}\label{sn13dk_identif}

Pre-explosion images of the SN site in the {\sl HST\/} archive are from the Wide Field Camera 3 (WFC3) Ultraviolet-Visible (UVIS) Channel  
(pixel scale of $0{\farcs}04$ pix$^{-1}$) and the Advanced Camera for Surveys (ACS) Wide Field Channel (WFC) (pixel scale of $\sim 0{\farcs}05$ pix$^{-1}$). The UVIS data are in bands $F336W$ ({\it $\sim U$}; 5530~s), $F555W$ ({\it $\sim V$}; 1032~s), $F625W$ ({\it $\sim r$}; 1016~s), and $F814W$ ({\it $\sim I$}; 1032~s), and were taken on 2011 January 22 and 28 UT ({\sl HST\/} ID proposal: 11577; PI: B. Whitmore; e.g. see \citealt{peacock13}). The ACS data are in band $F435W$ ({\it $\sim B$}; 1096~s) ({\sl HST\/} ID proposal: 10188; PI: B. Whitmore; e.g. see \citealt{whitmore10}). We worked with drizzled images downloaded from the Hubble Legacy Archive\footnote[11]{\it http://hla.stsci.edu/hlaview.html}. These images were resampled to a uniform grid to correct for geometric distortions. To identify the SN position in the {\sl HST\/} images, we used a $K_s$-filter NaCo ($\sim 0{\farcs}05$ pix$^{-1}$, $56\arcsec \times 56\arcsec$ field of view; 690~s total on-source) image obtained on 2013 July 02.98 UT on the VLT at Cerro Paranal (European Southern Observatory). Each of the 10 frames was sky subtracted using off-source sky frames bracketing the on-source exposures; the sky-subtracted frames were then shifted and added using {\sc IRAF}\footnote[12]{{\sc IRAF} (Image Reduction and Analysis Facility) is distributed by the National Optical Astronomy Observatories, which are operated by the AURA, Inc., under cooperative agreement with the National Science Foundation (NSF).} tasks.

\begin{figure*}
\centering
\includegraphics[width=2\columnwidth]{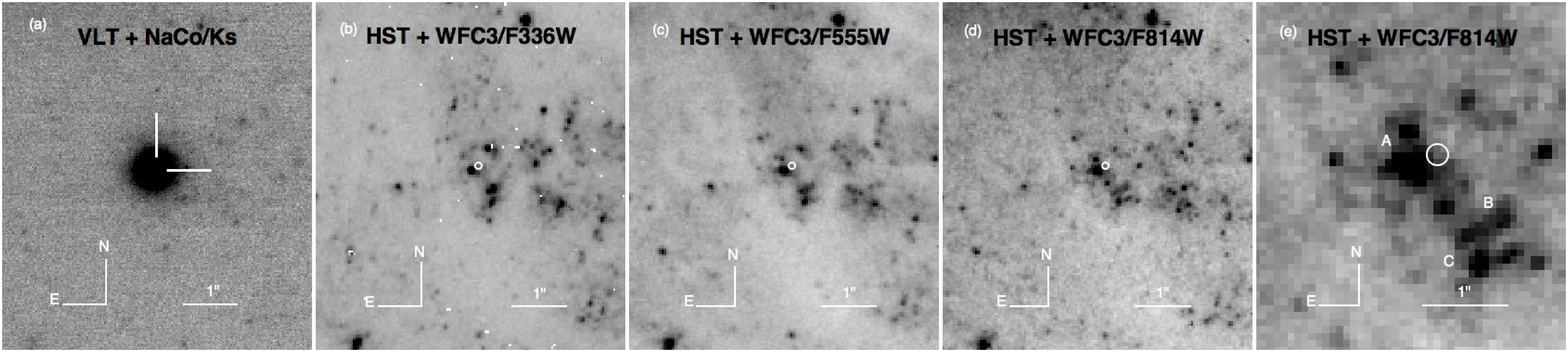}
\caption{Subsections of the post-explosion adaptive optics {\it $K_{\rm s}$} image of SN~2013dk with VLT + NaCo ({\it panel a}), and the pre-explosion {\sl HST\/} + WFC3 images in $F336W$ ({\it panel b}), $F555W$ ({\it panel c}), and $F814W$ ({\it panel d}) of NGC~4038. {\it Panel e} is a magnification of the {\sl HST\/} + WFC3 image in $F814W$. The positions of the SN is indicated, as well as three neighbouring sources of SN~2013dk, ``A", ``B" and ``C". 
}
\label{fig_progenitor}
\end{figure*}

We achieved high-precision relative astrometry between the {\sl HST\/} pre-explosion drizzled full mosaic images, and the NaCo post-explosion image, by geometrically transforming the former to the latter. We registered 30 point-like sources in common between the two datasets and measured their pixel coordinates with the {\sc IRAF} tasks {\it imexamine} and {\it daofind\/} (within {\sc IRAF/DAOPHOT}). Then, using {\it geomap}, and {\it geoxytran} (always {\sc IRAF} tasks), we carried out a geometrical transformation between the sets of coordinates. The position of the SN (and its uncertainty) is derived by averaging the measurements from the two {\sc IRAF} centroiding methods above mentioned. As a result, the SN location lies on the edge of a bright point-like source (labelled as ``A'' in Figure \ref{fig_progenitor}) visible in all bands. More precisely, it is located approximately at $0{\farcs}15$W and $0{\farcs}08$N from the center of Source ``A" (average values of the offsets in the different bands). The total estimated uncertainty in the astrometry is given in Table \ref{table_sources}. These values were calculated as a quadrature sum of the uncertainties in the SN position, and the {\it rms} uncertainty in the geometric transformation.

Source ``A" is one of detected sources near the pre-explosion error circle suggested by \citet{eliasrosa13}, however, the difference between the locations of ``A" and the SN location is now greater than $10\sigma$ (see Table \ref{table_sources}). Hence Source ``A"  can not be a single stellar precursor to SN 2013dk; however we will consider the possibility that it is a complex or cluster with which the progenitor was associated. Two other ``progenitor candidates" were identified in the preliminary analysis of \citet{eliasrosa13} (``B" and ``C" in Figure \ref{fig_progenitor}), and can now discounted as being related to SN~2013dk. 


\begin{table*}
\centering
\caption{Position and magnitudes of the Source ``A" and SN~2013dk. }
\label{table_sources}
\begin{tabular}{@{}lcccccc@{}}
\hline
 & Position$^{(a)}$ & $F336W$ & $F435W$ & $F555W$ & $F625W$ & $F814W$\\
 &  & (mag) & (mag) & (mag) & (mag) & (mag)\\
\hline%
Source ``A"  & $0{\farcs}15$E, $0{\farcs}08$S & 23.13(0.25) & 23.41(0.03) & 23.45(0.05) & 22.83(0.08) & 22.08(0.07) \\
Upper limit  & & $\gtrsim$27.2 & $\gtrsim$26.5 & $\gtrsim$26.8 & $\gtrsim$26.4 & $\gtrsim$24.9 \\
\hline
\hline
 & &  ($\alpha/\delta$) & ($\alpha/\delta$) & ($\alpha/\delta$) &  ($\alpha/\delta$) &  ($\alpha/\delta$)\\
\hline
Total uncertainty (mas) & & 6/6 & 9/8 & 6/6 & 6/6 & 5/6 \\
Diff. position SN/candidate ``A" (mas) & & 133/85 & 137/88 & 146/83 & 148/87 & 153/83\\
\hline
\end{tabular}
\begin{flushleft}
$^{(a)}$ Position of Source ``A" with respect to the SN location. 
\end{flushleft}
\end{table*}

%
%
\vspace{-0.5cm}
\section{Progenitor analysis}\label{sn13dk_natureprog}
We obtained photometry of the Source ``A" close to the SN position in the pre-explosion {\sl HST\/} images using the package DOLPHOT\footnote[13]{DOLPHOT is a stellar photometry package that was adapted from HSTphot for general use \citep{dolphin00}. We used the WFC3 and ACS modules of the v2.0, updated 2013 March. {\it http://americano.dolphinsim.com/dolphot/}. Note that DOLPHOT is not designed to run on the drizzled image mosaics.}. No progenitor star is visible at the exact SN position in any of these images. For all filters, it was necessary to measure small dithered offsets between exposures with respect to one fiducial image before running DOLPHOT. We rejected those images with such a low signal-to-noise ($< 5$) that Source ``A" was hardly detectable. DOLPHOT does not yet support  transformations from WFC3 flight-system magnitudes to the corresponding Johnson-Cousins magnitudes, hence we quote magnitudes in VEGAMAGS flight-system (Table \ref{table_sources}). 

DOLPHOT also reports a set of parameters to interpret the nature of the source, such as the ``object type'' flag, which in our case was ``1'' for all cases, meaning that the source is likely stellar. However, another relevant parameter is the ``sharpness", which in our case varies between $-0.18$ and $0.13$, indicating a likely good star, although negative values could also be signs of a broad source (\citealt{dolphin00}). In addition, the full width half maximum of the stellar point-spread function (PSF) is on average $\sim 2.6$ pixels, which at a distance of 22.3 Mpc corresponds to $\sim 11$ pc, which is quite large. The residual images, after PSF subtraction by DOLPHOT, are quite clean around the position of source ``A" in almost all of the images. We note there is possibly a small residual in the F336W and F555W images,  which could result from a poor subtraction due to the low signal-to-noise; however, we can not rule out the possibility of an additional contaminating source. Hence even if Source ``A'' seems to be a point-like source, the spatial size of the PSF at this distance means that it could be also a compact cluster. In the following, we consider the three possible interpretations of Source ``A'' and the progenitor of SN 2013dk:

\begin{itemize}
\vspace{-0.7mm}
\item {\bf Cluster}. While Source ``A'' is not clearly extended, its absolute magnitude is $M_{F555W} = -9.02 \pm 0.28$, brighter than the threshold of $M_V < -8.6$ mag which \cite{bastian05} and \cite{crockett08} suggest for compact star clusters {(assuming $F555W \sim V$)}. The photometry for Source ``A'' was compared with template cluster SEDs, calculated with Starburst99 \citep{leitherer99}, using our own minimisation and fitting code (Maund, 2013, in prep.), with a single instantaneous episode of star formation. The best fit solutions were consistent with moderate reddening [$E(B-V) \sim 0.2$] and ages in the range $15-22$ Myr.  The corresponding $\chi^{2}$ value of the best-fit solution was 39.1, and Figure \ref{fig_cluster} shows that the fit is not  excellent. If the progenitor of SN~2013dk was a member, {\em and} if we could assume that it was born coeval with the cluster,  then this would imply a turn-off mass of $10-15$ M$_{\odot}$ \citep[from the age-mass relations in][]{eldridge04,smartt09}.  This would favour a low mass helium star in an interacting binary as the progenitor and not a massive WR star \citep{podsiadlowski92,eldridge13}.
However, we caution that SN~2013dk is not coincident with Source ``A'', and hence it may not even be related. Moreover, the mean half-light or ``effective" radius found for young clusters in the Antennae Galaxies by \cite{whitmore99} was $4 \pm 1$ pc (although clusters can have also radiis $\sim 10$ pc; \citealt{larsen04}), and the projected distance from the SN location to the Source ``A" is between $9$ and $16$ pc. The strongest statement we can make is that we can find no evidence of a very young stellar population in local vicinity of SN~2013dk. 

\vspace{-0.7mm}
\item {\bf Single star}. As Source ``A'' is not inconsistent with a point-like source, we can not rule out the possibility that it is a single star. In this scenario, Source ``A'' would be unrelated to SN 2013dk. The photometry of Source ``A'' was compared with the single star SED models of \cite{castelli04}, again using our own fitting code.  The fit was unsatisfactory, yielding $\chi^{2}=83.9$ and implying a an extinguished [$E(B-V)=0.5$] hot star with $T\sim 14\,000K$, and the corresponding luminosity $\log (L/L_{\odot})=6.2$ would be extremely high for a single star.


\begin{figure}
\centering
{\includegraphics[width=1\columnwidth,angle=0]{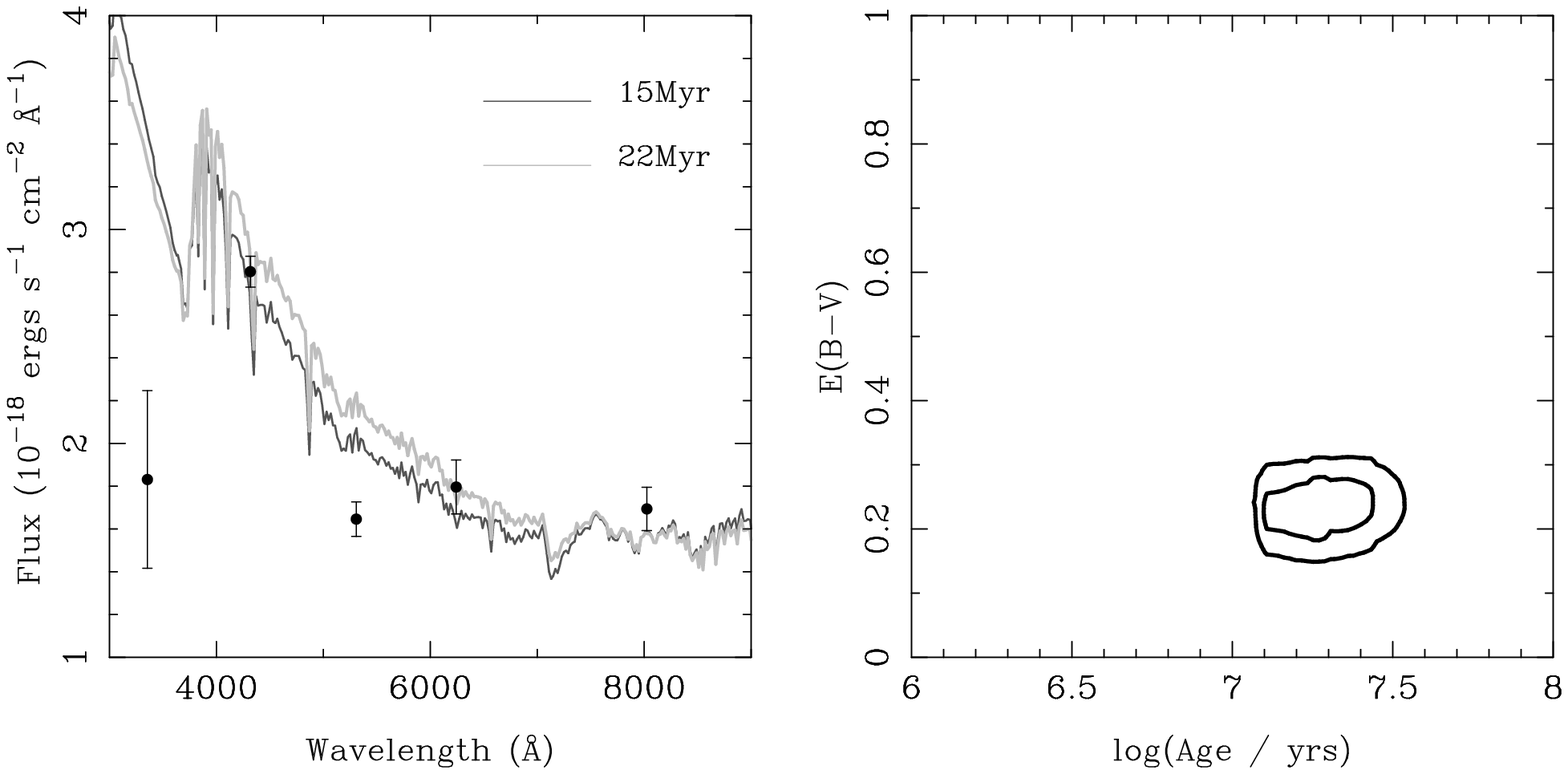}}
\caption{
{\bf Left:} Observed spectral energy distribution for Source ``A''. Overlaid are Starburst 99 model SEDs for clusters, with a single burst 
of star formation, with ages $15$ Myr (heavy line) and $22$ Myr (grey line). {\bf Right:} The joint posterior distribution for the reddening and 
cluster age for Source ``A''. The contours contain 68 and 95\% of the probability.}
\label{fig_cluster}
\end{figure}


\vspace{-0.7mm}
\item {\bf Progenitor not detected}. As there is no detection of a single star at the position of SN~2013dk, we can set upper limits to its magnitudes in each filter. If Source ``A'' is a single star, then these upper limits are the best restrictions we can place, and compare to the results in \cite{eldridge13}. We added artificial stars at the precise SN position (see \S~\ref{sn13dk_identif}) with the PSF created within DOLPHOT. The input brightness of the artificial star was in all cases fainter than  Source ``A". Thus, we reran DOLPHOT adding specific parameters for the insertion and measurement of the synthetic stars. The resulting upper limits are presented in Table \ref{table_sources}. Note that this methods works for both WFC3 and ACS {\sl HST\/} instruments. Adopting a distance modulus $\mu = 31.74 \pm 0.27$ mag, and $A_{V,\rm tot} = 0.73 \pm 0.06$ mag as the extinction toward SN~2013dk, we estimate a limit in the absolute magnitude of an undetected progenitor, $M_{F555W} \gtrsim -5.7$ mag. The magnitude limits can not rule out massive WC or WO stars as progenitors (WR stars with strong carbon or oxygen emission-lines), since in our Galaxy (see e.g. \citealt{sander12}) and  the Large Magellanic Cloud (e.g. \citealt{massey02}) such stars have magnitudes between -2 and -8 mag. Thus, the progenitor of SN~2013dk joins the list of undetected supernova Ib/c progenitors as summarised and described in \cite{eldridge13}. However, given also the large extinction suffered by this SN, we can not rule out that the progenitor could be hidden by  strong foreground dust that we have not accounted for. These magnitude limits alone do not allow us to distinguish robustly between high mass WR stars and lower mass helium star progenitors in interacting binaries \cite[see discussion in][]{eldridge13}, but it is useful to add these constraints to that sample. Including SN~2013dk in the same methodology as in \cite{eldridge13} reduces the probability that Ib/c SNe come exclusively from the massive WR population to around 12 per cent. 
\end{itemize}

In summary we cannot definitively determine that Source ``A'', the closest bright source to SN~2013dk in {\sl HST\/} images, is a cluster or single point source. Although it is unresolved, we favour it being a compact cluster given its bright absolute magnitude and more extended wings. We find a reasonable, but not excellent, fit to the cluster SED giving quite an age of $15-22$ Myrs. If the progenitor of SN~2013dk had been a cluster member, and coeval with cluster formation then this would imply a progenitor mass in the range $10-15$ M$_{\odot}$. However even if Source ``A'' is a cluster, SN~2013dk is not spatially coincident with it, nor is it within the half light radius. This limits what we can definitely determine about the progenitor nature. Future observations of the SN field with {\sl HST\/} once we are confident the SN has faded might allow difference imaging to verify if any source in the field has vanished, or at least has significantly weakened \citep[e.g. see][]{maund09,maund13}.

%

\vspace{-0.5cm}
\section*{Acknowledgments}

N.E.R. is supported by the MICINN grant AYA2011-24704/ESP, by the ESF EUROCORES Program EuroGENESIS (MICINN grant EUI2009-04170), by SGR grants of the Generalitat de Catalunya, and by EU-FEDER funds. The research of J.R.M. is supported through a Royal Society University Research Fellowship. A.P., S.B. are partially supported by the PRIN-INAF 2011 with the project ``Transient Universe: from ESO Large to PESSTO". The research leading to these results has received funding from the European Research Council under the European Union's Seventh Framework Programme (FP7/2007-2013)/ERC Grant agreement n$^{\rm o}$ [291222]  (PI : S. J. Smartt). K.T. and G.P. acknowledges support from Millennium Center for Supernova Science (P10-064-F), with input from Fondo de Innovaci\'on para la Competitividad, del Ministerio de Economia, Fomento y Turismo de Chile. K.T. is supported by proyecto Gemini-Conicyt 32110024. G.P. acknowledges partial support by proyecto interno UNAB DI-303-13/R. N.E.R. thanks J. J. Eldridge for making available the binary stellar models. This research is based in part on observations made with the NASA/ESA {\it Hubble Space Telescope}, and obtained from the Hubble Legacy Archive, which is a collaboration between the Space Telescope Science |Institute (STScI/NASA), the Space Telescope European Coordinating Facility (ST-ECF/ESA) and the Canadian Astronomy Data Centre (CADC/NRC/CSA). This work has made use of the NASA/IPAC Extragalactic Database (NED), which is operated by the Jet Propulsion Laboratory, California Institute of Technology, under contract with NASA.

 \vspace{-0.5cm}
\bibliography{bibtex_sn13dkprog}{}
\bibliographystyle{mn2e}

\bsp

\label{lastpage}

\end{document}